\documentclass[nofigure]{pasj00-ngvlaj}
\usepackage{graphicx}

\usepackage{times}
\usepackage{threeparttable}
\usepackage{url}

\begin{document}

\title{Predictions of dust continuum observations of circumplanetary disks with ngVLA: A case study of PDS~70~c}
\author{Yuhito Shibaike\altaffilmark{1,2}, Takahiro Ueda\altaffilmark{3}, Misato Fukagawa\altaffilmark{2}}
\altaffiltext{1}{%
   Graduate School of Science and Engineering, Kagoshima University, 1--21--35 Korimoto, Kagoshima City, Kagoshima 890--0065, Japan}
   \altaffiltext{2}{%
   National Astronomical Observatory of Japan, 2--21--1 Osawa, Mitaka-shi, Tokyo 181--8588, Japan}
   \altaffiltext{3}{%
   Radio \& Geoastronomy Division, Harvard-Smithsonian Center for Astrophysics, 60 Garden Street, Cambridge, MA 02138, USA}
\email{yuhito.shibaike@sci.kagoshima-u.ac.jp}

\maketitle

\begin{abstract}
A gas giant forms a small gas disk called a ``circumplanetary disk (CPD)'' around the planet during its gas accretion process. The small gas disk contains dust particles like those in a protoplanetary disk, and these particles could be the building material of large moons. A young T Tauri star PDS~70 has two gas accreting planets, and continuum emission from one of the forming planets, PDS~70~c, has been detected by ALMA Bands~6 and 7, which is considered as the dust thermal emission from its CPD. We reproduce the emission with both bands and predict how the dust emission will be observed by ngVLA by expanding the range of the wavelength from submillimeter to centimeter. We find that the flux density of the dust thermal emission can be detected with ngVLA at Band 6 (3~mm) and probably with Band 5 (7~mm) as well. We also find that the size and shape of the CPD can be constrained by observations of ngVLA Band 6 with reasonable observation time.
\vspace{15mm}

\end{abstract}

\section{Introduction} \label{sec:introduction}
A gas giant forms a small gas disk around its body during the gas accretion process. The small disk is called a ``circumplanetary disk (CPD)'', which is considered as the birth place of large moons of the planet. Recently, continuum (sub-)millimeter emission from a gas accreting planet, PDS~70~c, has been detected by Atacama Large Millimeter/submillimeter Array (ALMA) at Bands 6 and 7. It is considered as the thermal emission from the dust in the CPD around the planet \citep{ise19,ben21,fas25}, since it is compact emission. PDS~70 has another gas accreting planet, PDS~70~b, sharing one large gap with the planet c. However, the planet b has diffuse emission, which is not considered to originate from a CPD, and so only the emission from PDS~70~c has been widely accepted as detection of a CPD. Detection of a CPD by ALMA remains difficult due to its faint intensity and the overlap with the emission from its parental protoplanetary disk \citep{and21}.

\citet{shi24} modeled the flux density of the dust thermal emission from the entire CPD of PDS~70~c and obtained constraints on the planet mass and the gas accretion rate by comparing the model prediction with the peak emission of PDS~70~c at ALMA Band~7 ($\lambda=855~\mu{\rm m}$), $86\pm16~\mu{\rm Jy~beam}^{-1}$ \citep{ben21}. After that, \citet{fas25} reported a detection of the emission from PDS~70~c at ALMA Band~6 with peak emission of $54\pm17~\mu{\rm Jy~beam}^{-1}$ ($\lambda=1.36~{\rm mm}$) and $54\pm14~\mu{\rm Jy~beam}^{-1}$ ($\lambda=1.15~{\rm mm}$), where the spectral index between ALMA Bands 6 and 7 is about $\alpha=2$. Although these observations only provide information about the total flux density due to insufficient spatial resolution, the Next Generation Very Large Array (ngVLA) has the potential to spatially resolve the CPD emission. In this memo, we calculate the dust thermal emission from the CPD of PDS~70~c with various wavelengths, reproduce the observations of ALMA Bands~6 and 7, and predict the future observations by ngVLA.

Note that \citet{dom25} reported detections of the continuum emission from PDS~70~c at ALMA Bands~3 and 4 and a non-detection at Band~9 very recently, but this work does not include them in the discussion, which is a future work.

\section{Methods} \label{sec:methods}
We use a model provided by \citet{shi24} for the calculation of the dust evolution in the CPD and the thermal emission from the dust. In this model, the steady radial distribution of the gas disk and the mid-plane temperature are first calculated. The dust evolution in the disk is then calculated, considering the growth of the particles with their mutual collision, fragmentation, and radial drift. Finally, the calculation of the intensity of the dust thermal emission at each radial distance from the central planet in any wavelength is carried out, considering the radial distribution of the dust temperature (assumed to be the same as the gas temperature), dust size, and dust surface density.

As shown in Table \ref{tab:properties}, we fix the planet properties according to \citet{shi24} and its references. PDS~70 is a young T Tauri star with the spectral type, mass, and age of K7, $M_{*}=0.76~M_{\odot}$, and $5.4~{\rm Myr}$ old, respectively. It is located at $d=113.43~{\rm pc}$ in the Upper Centaurus Lupus association \citep{gaia18,mul18}. The two gas accreting planets, PDS~70~b and c, have been found at the semi-major axes of $a_{\rm p}=20.6$ and $34.5~{\rm au}$, respectively, and the planets share a large gap in a pre-transitional disk with inclination of $i=51.7^{\circ}$ \citep{mul18,kep19,haf19}. We assume that the inclination (and the position angle) of the CPD of PDS~70~c is the same as that of the pre-transitional disk.

According to \citet{shi24}, the observed flux density of the continuum observations by ALMA Band 7 can be reproduced when the planet mass, gas accretion rate, dust-to-gas mass ratio in the gas inflow onto the CPD, and strength of turbulence in the CPD are $M_{\rm p}=10~M_{\rm J}$, $\dot{M}_{\rm g}=0.2~M_{\rm J}~{\rm Myr^{-1}}$, $x=0.01$, and $\alpha_{\rm CPD}=10^{-4}$, respectively (Tables \ref{tab:properties} and \ref{tab:model}). Here, it is assumed that the entire disk emission is encompassed within a single beam. The (minus) power-low index of the size frequency distribution (SFD) of dust is assumed to be $q=3.5$. The critical fragmentation velocity of the rocky and icy particles are assumed as $v_{\rm ice}=50~{\rm m~s}^{-1}$ and $v_{\rm rock}=5~{\rm m~s}^{-1}$, respectively. The planet radius and effective temperature are estimated from fitting of spectral energy distributions (SEDs) of near-infrared observations as $R_{\rm p}=2.0~R_{\rm J}$ and $T_{\rm p,eff}=1051~{\rm K}$, respectively, by \citet{wan21}. The midplane temperature of the parental disk at $a_{\rm p}=34.5~{\rm au}$, $T_{\rm PPD}=32~{\rm K}$, is estimated from a set of observations of CO isotopologue and HCO$^{+}$ by \citet{law24}.

However, very recent observations with ALMA Band~6 show that the spectral index between ALMA Bands~6 and 7 is about $\alpha=2$ \citep{fas25}, which is smaller than the model prediction using the parameter sets provided by \citet{shi24}. In this memo, we consider a case which can also reproduce the observations of ALMA Band~6. The small spectral index suggests that the disk is optically thick with the wavelength between ALMA Bands 6 and 7. We then assume $x=0.05$, $\alpha_{\rm CPD}=10^{-5}$, and $q=2.5$ to make the disk optically thick (Table \ref{tab:model}). This relatively high $x$ value could be an extreme assumption, because previous local hydrodynamical simulations suggest that the inflow to CPDs should be dust poor \citep{tan12,mae24}. However, recent JWST/NIRCam observations find a dust accretion stream bridging the outer ring of the pre-transitional disk to the vicinity of PDS~70~c \citep{chr24}, which is also predicted by global hydrodynamical simulations as the so-called ``meridional circulation'' \citep{szu22}. The weak $\alpha_{\rm CPD}$ and small $q=2.5$ could also be achieved in circumplanetary and/or protoplanetary disks \citep{dul18,shi24}. We then assume $M_{\rm p}=8~M_{\rm J}$ to adjust the flux density at ALMA Bands~6 and 7 observed by \citet{fas25}. We also assume $v_{\rm ice}=v_{\rm rock}=1~{\rm m~s}^{-1}$, considering that recent observations of protoplanetary disks suggest fragile particles \citep{oku19,ued24}. The dust-to-gas surface density ratio used for the calculation of the disk temperature is fixed as $Z_{\Sigma,{\rm est}}=10^{-4}$ for the simplification (see Appendix B of \citet{shi25}).

\begin{table}[htbp]
\centering
\small
\caption{Properties of PDS~70 and PDS~70~c}
\label{tab:properties}
\begin{tabular}{lll}
\hline
\hline
PDS~70 & & \\
\hline
Host star mass & $M_{\rm *}$ & $0.76~M_{\odot}$ \\
Distance from Earth & $d$ & $113.43~{\rm pc}$ \\
Inclination of PPD (CPD) & $i$ & $51.7^{\circ}$ \\
\hline
PDS~70~c & & \\
\hline
Semimajor axis & $a_{\rm p}$ & $34.5~{\rm au}$ \\
Radius & $R_{\rm p}$ & $2.0~R_{\rm J}$ \\
Effective temperature & $T_{\rm p,eff}$ & $1051~{\rm K}$ \\
Temperature of PPD & $T_{\rm PPD}$ & $32~{\rm K}$ \\
\hline
\end{tabular}
\end{table}

\begin{table}[htbp]
\centering
\small
\caption{Model assumptions}
\label{tab:model}
\begin{tabular}{lll}
\hline
\hline
This work & & \\
\hline
Planet mass & $M_{\rm p}$ & $8~M_{\rm J}$ \\
Gas accretion rate & $\dot{M}_{\rm g}$ & $0.2~M_{\rm J}~{\rm Myr^{-1}}$ \\
Dust-to-gas mass ratio in inflow & $x$ & $0.05$ \\
Strength of turbulence in CPD & $\alpha_{\rm CPD}$ & $10^{-5}$ \\
Power-low index of dust SFD & $q$ & $2.5$ \\
Fragmentation velocity (ice) & $v_{\rm ice}$ & $1~{\rm m~s}^{-1}$ \\
Fragmentation velocity (rock) & $v_{\rm rock}$ & $1~{\rm m~s}^{-1}$ \\
\hline
\citet{shi24} & & \\
\hline
Planet mass & $M_{\rm p}$ & $10~M_{\rm J}$ \\
Gas accretion rate & $\dot{M}_{\rm g}$ & $0.2~M_{\rm J}~{\rm Myr^{-1}}$ \\
Dust-to-gas mass ratio in inflow & $x$ & $0.01$ \\
Strength of turbulence in CPD & $\alpha_{\rm CPD}$ & $10^{-4}$ \\
Power-low index of dust SFD & $q$ & $3.5$ \\
Fragmentation velocity (ice) & $v_{\rm ice}$ & $50~{\rm m~s}^{-1}$ \\
Fragmentation velocity (rock) & $v_{\rm rock}$ & $5~{\rm m~s}^{-1}$ \\
\hline
\end{tabular}
\end{table}

\section{Results} \label{sec:results}
\subsection{Detectability of the CPD of PDS~70~c} \label{sec:detectability}
\begin{figure*}[htbp]
\centering
\includegraphics[width=0.8\linewidth]{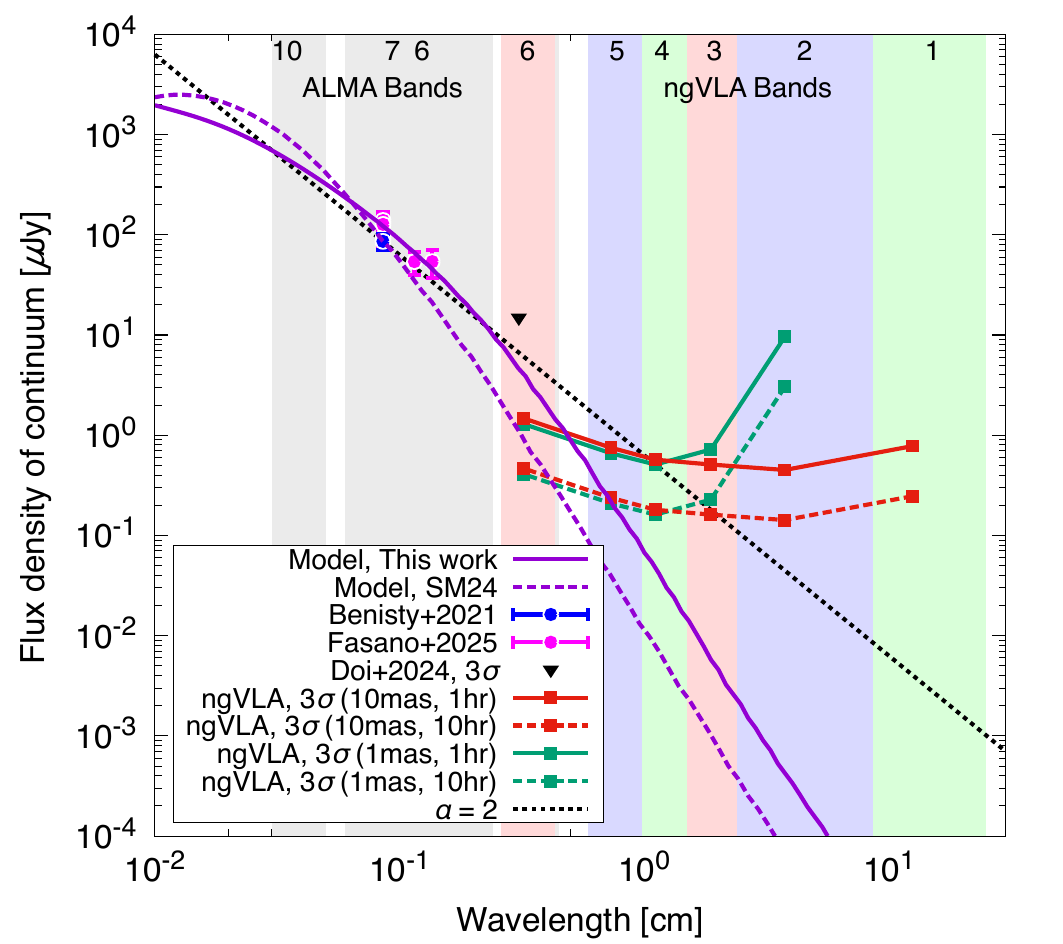}
\caption{Detectability of the dust thermal emission from the CPD of PDS~70~c with ngVLA. Purple curves are the model predictions of the total flux density of the dust thermal emission from the CPD; the solid and dashed curves are the cases of this work and \citep{shi24}, respectively (see Table \ref{tab:model}). Blue circle represents the flux density observed by ALMA Band 7 ($\lambda=855~\mu{\rm m}$), $86\pm16~\mu{\rm Jy}$ \citep{ben21}. Magenta circles are the flux density recently observed by ALMA Band~6, $54\pm17~\mu{\rm Jy}$ ($\lambda=1.36~{\rm mm}$) and $54\pm14~\mu{\rm Jy}$ ($\lambda=1.15~{\rm mm}$), and by ALMA Band~7 ($\lambda=856~\mu{\rm m}$), $94\pm18~\mu{\rm Jy}$, $127\pm23~\mu{\rm Jy}$, and $139\pm28~\mu{\rm Jy}$, reported by \citet{fas25}. Black triangle represents the $3\sigma$ sensitivity of the previous non-detecting observations of ALMA Band~3 ($\lambda=3.07~{\rm mm}$), $14.82~\mu{\rm Jy}$ \citep{doi24}. Red solid curve is the $3\sigma$ sensitivity of ngVLA observations with $10~{\rm mas}$ spatial resolution and 1 hour on-source time. Dashed red curve is that with 10 hour on-source time (i.e., $\sqrt{10}$ times higher sensitivity). Green curves are those with $1~{\rm mas}$ spatial resolution. Black dotted line is the flux density with the spectral index $\alpha=2$ reproducing the observed value of ALMA Band 7 \citep{ben21}. Color and gray shaded regions represent the bands of ngVLA and ALMA, respectively.}
\label{fig:detectability}
\end{figure*}

First, we investigate the detectability of the CPD of PDS~70~c with the future ngVLA observations. We vary the wavelength for $10^{-2}~{\rm cm}\leq\lambda\leq10^{2}~{\rm cm}$ and compare the predicted flux density of the dust thermal emission from the CPD with the sensitivity of possible ngVLA observations. Figure~\ref{fig:detectability} shows that our model (solid purple curve) can reproduce the flux density observed by \citep{fas25} at both ALMA Bands~6 and 7 (magenta circles), where the curve is along with the line representing $\alpha=2$ between the two bands (black dotted line). As a result, the predicted flux density is higher than that predicted using the parameter sets of \citep{shi24} (dashed purple curve) in the longer wavelength than that of ALMA Band~7.

Figure~\ref{fig:detectability} also shows that the predicted flux density is higher than the $3\sigma$ sensitivity of ngVLA at Band~6 ($\lambda=3.22~{\rm mm}$) with $10~{\rm mas}$ or $1~{\rm mas}$ spatial resolutions and $1~{\rm hr}$ on-source time. On the other hand, with the assumptions of \citet{shi24}, the predicted flux density is higher than the $3\sigma$ sensitivity only with $10~{\rm hr}$ on-source time. At ngVLA Band~5 ($\lambda=7.31~{\rm mm}$), the flux density predicted by our model is about $3\sigma$ sensitivity with $10~{\rm hr}$ on-source time. Therefore, ngVLA will clearly detect the dust thermal emission from the CPD of PDS~70~c at Band~6 and will possibly detect it at Band~5. With the other ngVLA Bands, it would be difficult to detect the CPD. Figure \ref{fig:detectability} also shows that it is difficult to detect the CPD with ngVLA Bands~3 to 1, even if the spectral index is $\alpha=2$ at those wavelengths. Therefore, the sweet spot for observing CPDs with ngVLA lies in Bands~4 to 6.

\subsection{Possible images of the CPD of PDS~70~c} \label{sec:resolved}
\begin{figure*}[htbp]
\centering
\includegraphics[width=0.45\linewidth]{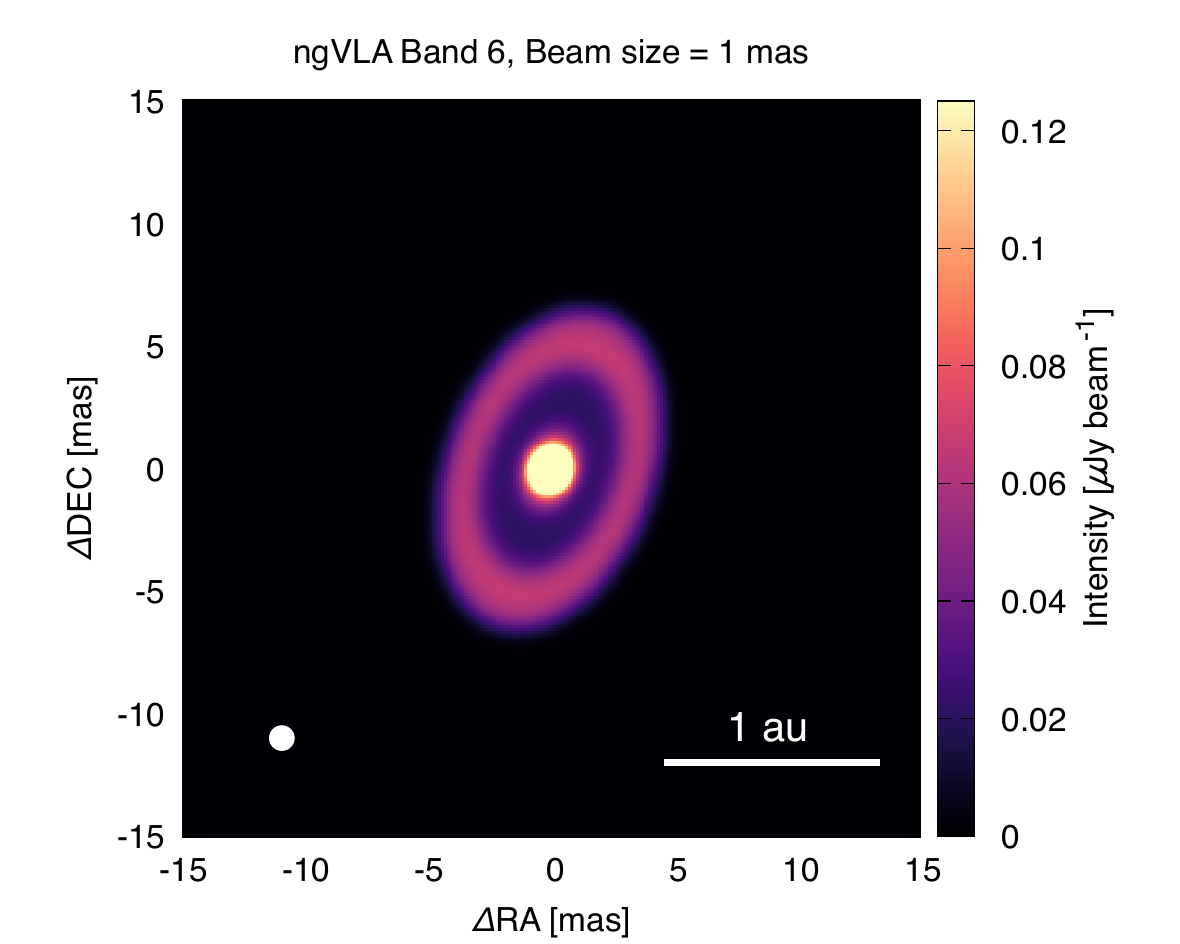}
\includegraphics[width=0.45\linewidth]{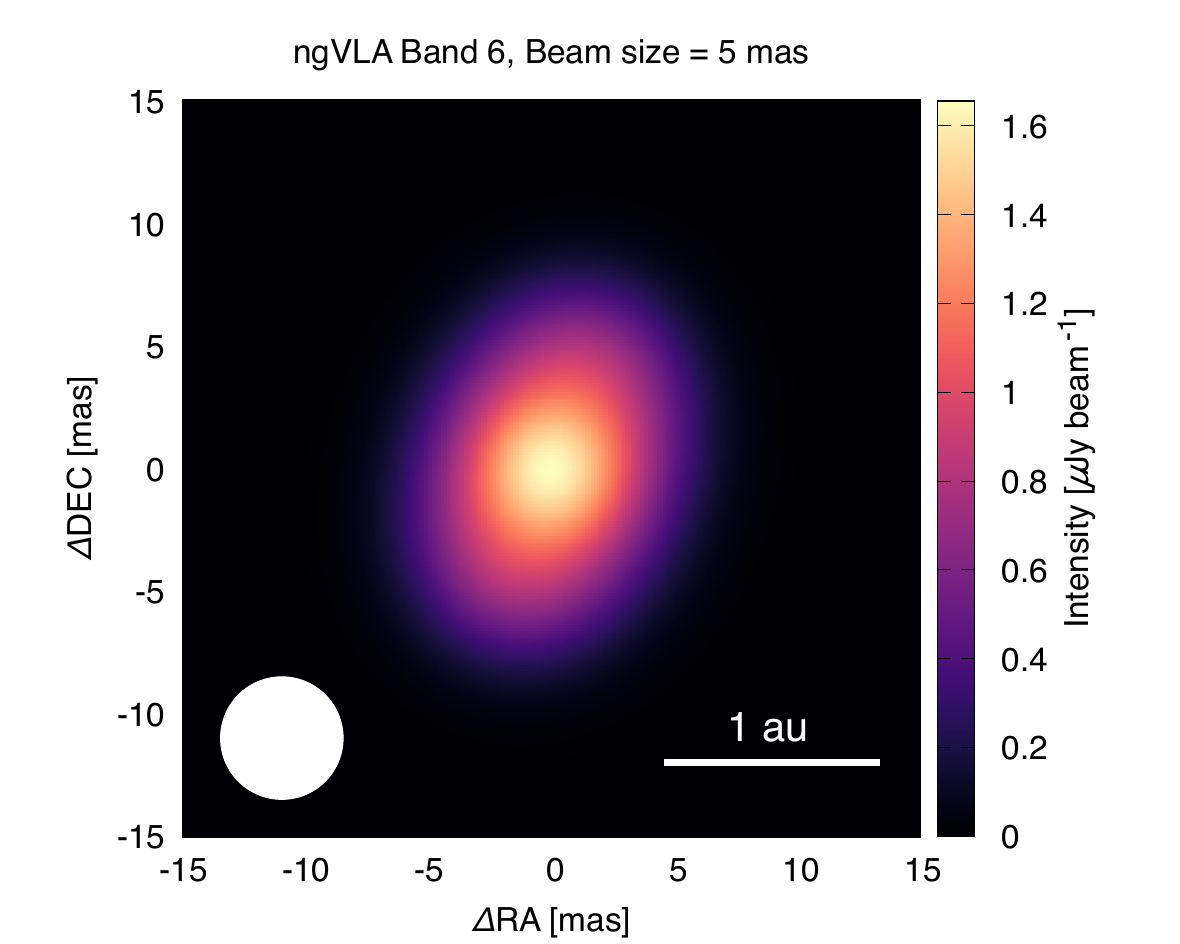}
\includegraphics[width=0.45\linewidth]{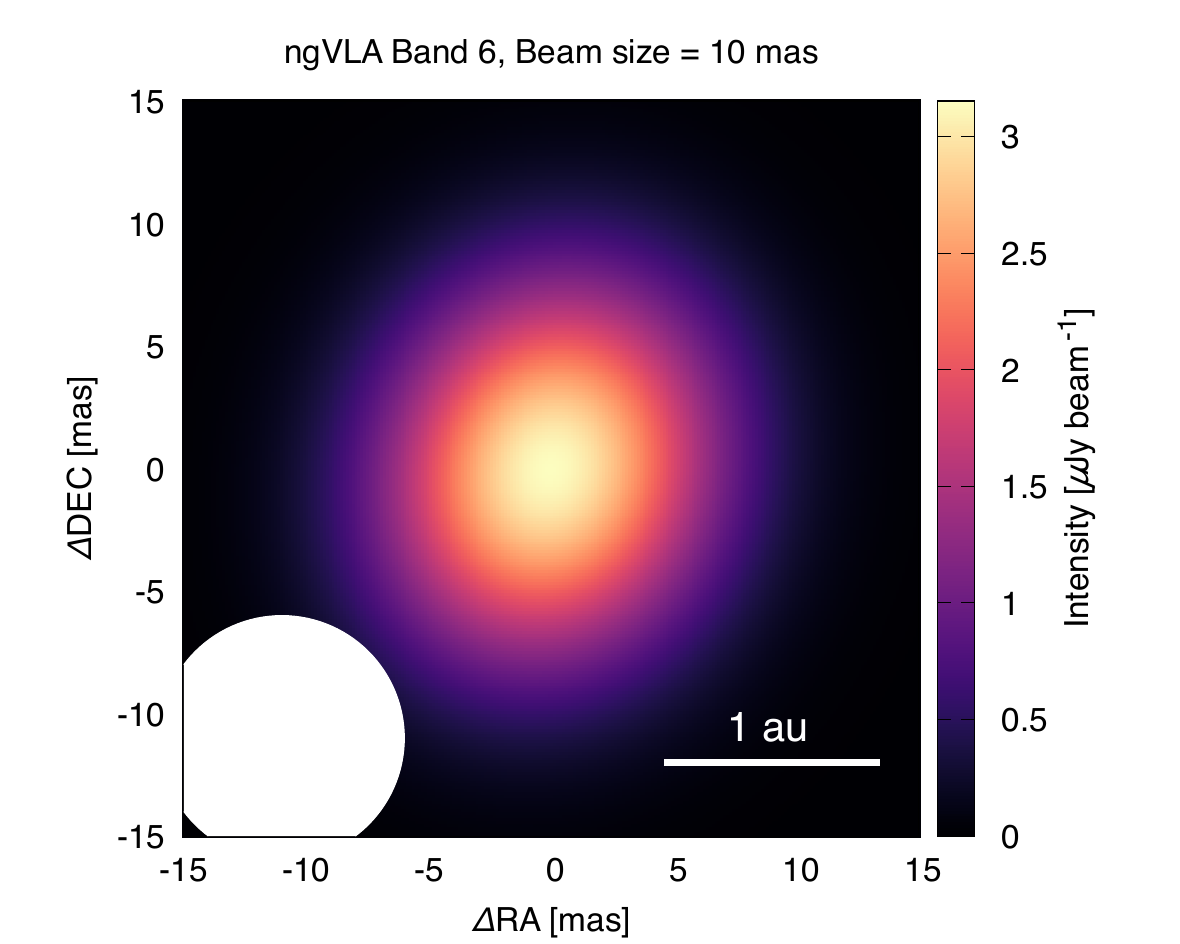}
\includegraphics[width=0.45\linewidth]{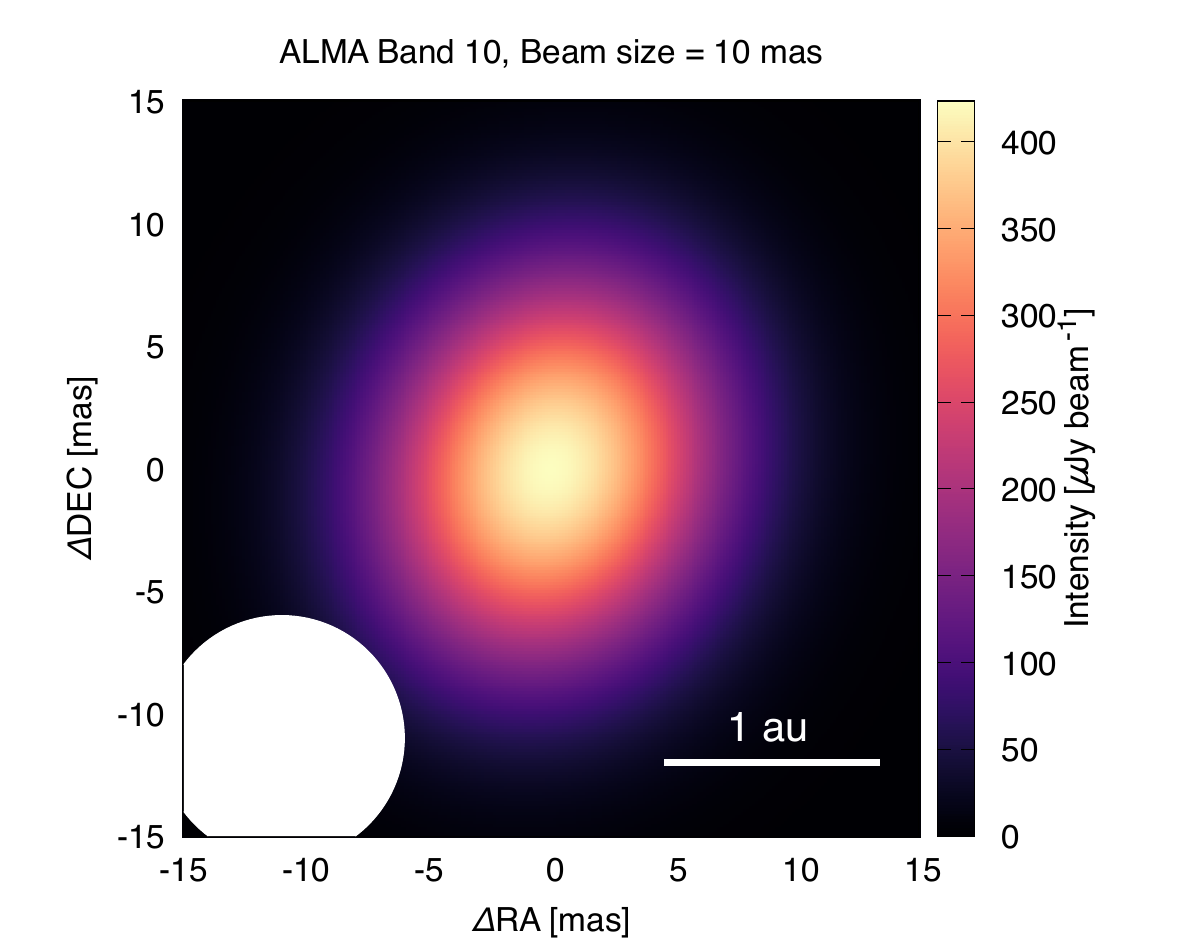}
\caption{Predicted intensity maps of the dust thermal emission from the CPD of PDS~70~c smoothed with different beam sizes. The upper left, upper right, and lower left panels represent the maps of the wavelength of ngVLA Band~6 ($\lambda=3.22~{\rm mm}$) with the beam sizes of 1 mas, 5 mas, and 10 mas, respectively. The lower right panel represents the map of ALMA Band~10 ($\lambda=345~\mu{\rm m}$) with the beam size of 10 mas, the highest spatial resolution by ALMA. The white circles at the left bottom corners represent the beam sizes. The maximum values of the color bars are set to be 1/5 of the maximum intensity for the upper left map and the maximum intensity for the other maps.}
\label{fig:images}
\end{figure*}

We then predict the possible images obtained by ngVLA and ALMA. Figure \ref{fig:images} shows the intensity maps of the CPD predicted by our model when the CPD is axially symmetric and has the same inclination and position angle as the parental pre-transitional disk. We investigate the maps with the shortest wavelength of ngVLA, Band 6 ($\lambda=3.22~{\rm mm}$), since the CPD can be detected at the band as we show in the previous section. We consider the cases with the beam sizes smaller than 10 mas, since the CPD can be distinguished from the outer ring of the pre-transitional disk with the ngVLA spatial resolutions \citep{ben21}.

The lower left panel shows that the size of the CPD can be roughly estimated by the observations with the beam size of 10 mas if the outer parts of the disk, with $\sim1~\mu{\rm Jy~beam}^{-1}$, are detected. This condition is reasonable, because the on-source time for $4\sigma$ detection will be about four hours with the sensitivity of $0.25~\mu{\rm Jy~beam}^{-1}$ (Table \ref{tab:ngVLA}; \citet{ngvla_performance_2021}). Although this spatial resolution can also be achieved by ALMA with Band~10 ($\lambda=345~\mu{\rm m}$) C-9 configuration, we find that it is difficult to detect the outer parts ($\sim200~\mu{\rm Jy~beam}^{-1}$; the lower right panel) with reasonable observation time; it takes longer than 200 hours to achieve the sensitivity by ALMA Band~10 (Table \ref{tab:ngVLA}).

The upper two panels represent the intensity maps smoothed with 1 mas (left) and 5 mas (right) beam sizes. The panels show that the shape of the CPD can be constrained from those spatial resolutions. The intensity of the outer parts of the disk is roughly $\sim0.5~\mu{\rm Jy~beam}^{-1}$ with 5 mas, and those parts will be detected with $5\sigma$ with reasonable on-source time, $\sim20~{\rm h}$ (Table \ref{tab:ngVLA}). This means that information of the inclination and/or position angle of the CPD can be obtained by ngVLA. Recent ALMA observations of SO line emission detected an outflow from a potential forming planet and/or its CPD inside a gap in TW~Hya disk, and its axis is likely tilted from the norm of the protoplanetary disk by $\sim50^{\circ}$ \citep{yos24}. If such an outflow is magnetic disk wind lunched from a CPD \citep{shi23}, the CPD could have a different inclination and/or position angle from those of the parental protoplanetary disk. The origin of the tilted axis of Uranus could also be related with a tilted axis of the potential CPD of Uranus \citep{rog20}. The understanding of the inclination and position angle of the CPD of PDS~70~c will provide important suggestions to such discussion beyond the CPD property itself.

The upper left panel shows that the substructures of the CPD could be resolved with 1 mas beam size. The faint ring like region and the inner bright region are shown in the intensity map. We find that the inner edge of the ring is the boundary of the Epstein and Stokes regimes of the dust particles in the CPD in our model; dust particles quickly drift inward and reduce their surface density and intensity in the Stokes regime. The intensity of the faint ring is only $\sim0.06~\mu{\rm Jy~beam}^{-1}$, and the on-source time to detect the ring with $3\sigma$ (i.e., with the sensitivity of $0.02~\mu{\rm Jy~beam}^{-1}$) is $\sim400~{\rm h}$ (Table \ref{tab:ngVLA}), which is not reasonable. However, if protoplanetary disks have substructures such as pressure bumps, dust particles pile-up there and make much brighter rings \citep{dul18}, which could also be the case in CPDs. Pressure bumps should be able to form in CPDs at the outer edges of the magnetic disk wind regions and/or at the inner edges of gaps formed by large moons \citep{shi19,shi23}. Therefore, such brighter rings will potentially be detected by ngVLA, but the investigation of them is beyond the scope of this memo.

\begin{table}[htbp]
\centering
\small
\caption{Integration time of the observations with ngVLA and ALMA}
\label{tab:ngVLA}
\begin{tabular}{llll}
\hline
\hline
Telescopes & Beam size & Sensitivity & Time \\
\hline
ngVLA (3.22~mm) & 1~mas & 0.02 $\mu{\rm Jy~beam}^{-1}$ & $\sim$400 h \\
& 5~mas & 0.1 $\mu{\rm Jy~beam}^{-1}$ & $\sim$20 h \\
& 10~mas & 0.25 $\mu{\rm Jy~beam}^{-1}$ & $\sim$4 h \\
\hline
ALMA ($345~\mu{\rm m}$) & 10~mas & 780 $\mu{\rm Jy~beam}^{-1}$ & 16 h \\
& & 200 $\mu{\rm Jy~beam}^{-1}$ & $>$200 h \\
\hline
\end{tabular}
\end{table}

\section{Conclusions} \label{sec:conclustion}
Young T Tauri star PDS~70 has two forming planets sharing a large gap in a pre-transitional disk. Compact continuum emission has been detected in the gap with ALMA Bands~6 and 7, which has been considered as the thermal emission from the dust in the CPD of PDS~70~c. We reproduce the flux density of the emission with both of the ALMA Bands~6 and 7 using a model of dust evolution and emission in a CPD provided by \citet{shi24} and predict the emission with the ngVLA bands by expanding the range of the wavelength.

We first found that the dust thermal emission from the CPD can be detected by ngVLA at Band~6 and probably at Band~5 as well, with on-source time shorter than 10 hours. We then predicted the intensity maps of the CPD at ngVLA Band 6 and showed that the size and shape of the CPD will probably be constrained by ngVLA observations with reasonable on-source time, about 20 hours. However, it is difficult to detect the potential substructures of the CPD unless there are substructures such as gas pressure bumps forming much brighter dust rings compared to our predictions with a smoothed CPD of PDS~70~c.

\section*{Acknowledgment}
We thank Satoshi Okuzumi for helpful discussion. This work was supported by JSPS KAKENHI Grant Numbers JP22H01274 and JP24K22907.

\bibliographystyle{aasjournal-revised}
\bibliography{ngvla}

\end{document}